\documentclass{aa}            % TWO COLUMNS

\usepackage{graphicx}
\usepackage{txfonts}
\begin{document} 
%%%%%%%%%%%%%%%%%%% TITLE PAGE %%%%%%%%%%%%%%%%%%%

% Title of the paper, and the short title which is used in the headers.
% Keep the title short and informative.
\title
{OJ~287: a new BH mass  estimate  of the  secondary%Short title, max. 45 characters
}

% The list of authors, and the short list which is used in the headers.
% If you need two or more lines of authors, add an extra line using \newauthor
\author{ Lev Titarchuk \inst{1} %, 2} 
\and
Elena Seifina\inst{2}%\fnmsep\thanks{Just to show the usage
\and
Chris Shrader\inst{3,4}
}

   \institute{    Dipartimento di Fisica, University di Ferrara, Via Saragat 1,
I-44122, Ferrara, Italy,  \email{titarchuk@fe.infn.it}, \
% LAPTh, Universite Savoie Mont Blanc, CNRS, B.P. 110, Annecy-le-Vieux
%F-74941, France, \
%LAPTh, CNRS, Annecy-le-Vieux F-74941, France
        \and
Lomonosov Moscow State University/Sternberg Astronomical Institute,
Universitetsky Prospect 13, Moscow, 119992, Russia, \email{seif@sai.msu.ru}\
        \and
NASA Goddard Space Flight Center, NASA, Astrophysics Science Division, Code 661, Greenbelt, MD
20771, USA, \email{Chris.R.Shrader@nasa.gov}\
        \and
Universities Space Research Association, 10211 Wincopin Cir, Suite 500, Columbia, MD 21044, USA
}

% These dates will be filled out by the publisher
%\date{Accepted by Astronomy \& Astrophysics,  February 11, 2023 }.
% Received YYY; in original form ZZZ}
   \date{Received, January 17, 2023,
        	accepted February 11, 2023
}

% Enter the current year, for the copyright statements etc.
%%\pubyear{2022}

% Don't change these lines
%%\begin{document}
%%\label{firstpage}
%%\pagerange{\pageref{firstpage}--\pageref{lastpage}}
%%\maketitle

%\begin{abstract}
  \abstract
{We presented a study of outburst activity  in the BL Lacertae source OJ~287, observed extensively with the X-ray  telescope (XRT) onboard Neil Gehrels Swift Observatory. We demonstrated  that  the results of our analysis of X-ray flaring activity   using the Swift/XRT data allow to refine the  key characteristics of the OJ~287 secondary (its nature and mass). We discovered that  the energy spectra in all spectral states can be fitted using the XSPEC Bulk Motion Comptonization (BMC) model.  As a result we found  that the X-ray photon index of the  BMC model,  $\Gamma$ correlates with the mass accretion rate, $\dot M$. We established that $\Gamma$ increases monotonically with $\dot M$ from the low-hard state, $\Gamma\sim 1.5$  to the high-soft state, $\Gamma\sim2.8$ and finally  saturates. The  index behavior was  similar to that  in a number of  black hole (BH) candidates in which we showed that   its  saturation  was an observational evidence of the presence of a BH. Based on this correlation, we applied a scaling method and determined  that  a secondary BH mass in OJ~287 is about   $\sim1.25\times10^8$  solar masses, using the well-studied  X-ray BH  binaries XTE~1550--564, H~1743--322, 4U~1630--47, GRS~1915+105  as well as extragalactic BHs ESO~243--49 and M101~ULX--1, as  reference sources. Also using the power spectrum analysis we inferred the size of the Compton cloud  $L_{CC}\sim 10^{13}$ cm where X-ray spectra were formed. Using this value of   $L_{CC}$ we confirmed that a BH mass of the secondary in  OJ~287  was of order of $10^8$ solar masses as we derived using the index, $\Gamma-$correlation (the scaling method)  with respect of the mass accretion rate.
%%\end{abstract}
} 

\titlerunning{OJ~287: Black Hole Mass Estimate of the  Secondary}
%{On evaluation of BH mass in OJ~287}

   \maketitle

%\begin{keywords}
%black hole physics---BL Lacertae objects: individual: OJ~287--galaxies: active%:radiation mechanisms: % non-thermal---physical data and processes
%keyword1 -- keyword2 -- keyword3
%\end{keywords}

%%%%%%%%%%%%%%%%%%%%%%%%%%%%%%%%%%%%%%%%%%%%%%%%%%

%%%%%%%%%%%%%%%%% BODY OF PAPER %%%%%%%%%%%%%%%%%%

\section{Introduction}

\begin{figure*}
\centering
\includegraphics[scale=0.50,angle=0]{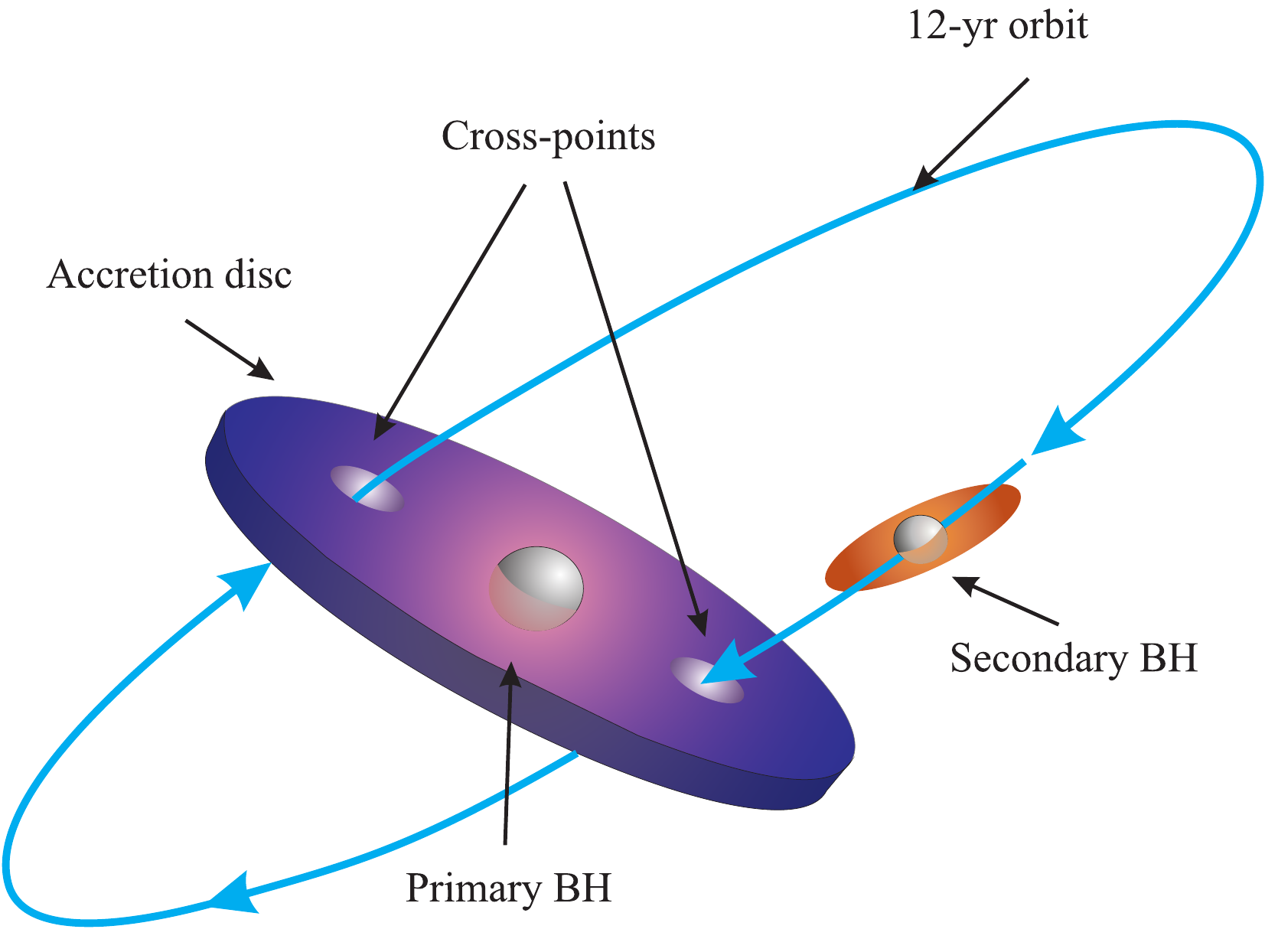}
\caption{A schematic view of OJ~287 model used in our analysis.
}
\label{picture}
\end{figure*}

Galaxy OJ~287 is  a BL Lacertae object (blazar) in the constellation Cancer and is located at a distance of five billion light years from Earth. As is known, blazars are associated with a supermassive black hole (SMBH), which collects the surrounding matter, dust and gas, forming an accretion disk. However, OJ~287 is interesting not only for this. At its center is not one, but two SMBHs \citep{Gomes22,Dou22}. These two black holes form an orbital pair located in the core of this galaxy. This pair is the only close binary system of two SMBHs known to date  \citep{val06,Laine20,Gomes22}. In turn, a large black hole (primary) has a mass equal to 18 billion solar masses, in fact equal to the mass of a small galaxy. The less massive black hole (secondary) weighs as much as 100 million solar masses. In addition, the secondary revolves around the primary (see Fig.~\ref{picture}), piercing/drilling through its accretion disk twice every 12 years \citep{Shi07,Dey19}. SMBHs binaries  are an amazing by-product of galaxy mergers in a hierarchical universe~\citep{Begelman80}. In the last stage of their orbital evolution, gravitational wave radiation provides the binary inspiral. 
Periodically varying radiation from active galactic nuclei has been proposed as a powerful tool for studying such binary systems~\citep{Chen20,Charisi16,Liu16,Zheng16,Graham15}. 

%%%%%%%%%%%%%%%%%%%%%%%%%%%%%%%%%%%%%%%%%%%%%%%
%
% FIGURE 1
%
%%%%%%%%%%%%%%%%%%%%%%%%%%%%%%%%%%%%%%%%%%%%%%%%%%%%

The goal of our paper is  to  check  a mass value of the secondary black hole (BH) in OJ~287 applying the scaling method for  a BH mass determination [\cite{st09}, hereafter ST09]  based on observations during source X-ray outbursts. Because OJ~287 exhibited a more or less strict 12-year cycle, several hypotheses have been proposed to explain the optical/X-ray periodic variability of the object \citep{sil88,Shi07}. In particular, according to \cite{val12},  hereafter V12, the periodicity is presumably due to the orbital rotation of the components in a BH binary, in which the secondary periodically perturbs the accretion disk around the primary. 

V12  presented results of the two X-ray observations of OJ~287 by XMM$-$Newton in 2005 April 12 and November 3--4. V12 claimed that the spectral energy distribution, the spectrum from radio to X-rays on April 12, 2005 followed nicely a synchrotron self-Compton (SSC) model [see \cite{cipr07}]. In contrast, the November 3--4 spectrum was quite different. The flux rose prominently in the optical/UV region, but in the radio or hard X-rays the flux remained at the pre-outburst level, making the single zone SSC model unlikely. However, V12 model of  the optical/UV outburst in the context of an interaction between the secondary and the primary accretion disk leading to a mass estimate of $M_{sec} \sim1.4\times10^8M_\odot$. V12 also formulated a question regarding the primary BH mass in order it could guarantee the stability of the primary accretion disc. They found that the minimum value of the primary mass $1.8 \times 10^{10} M_{\odot}$  is quite close to the BH mass determined from the orbit solution technique, 
$1.84 \times 10^{10} M_{\odot}$. 

To find evidence for the emission of the secondary, V12 needed to look at short time-scale variability in OJ~287. It has been found to be variable from 15 min time-scale upwards on many occasions [see for example, \cite{gup12}] and on one occasion the light curve has shown sinusoidal variations of the period of 228 min \citep{sag04}. If this variation is associated with the last stable orbit of a maximally rotating black hole, the mass of the black hole is $1.46\times 10^8 M_{\odot}$ \citep{gup12}, i.e. identical to the mass obtained from the orbit solution \citep{val10}.

 \cite{kom21}, hereafter KOM21 made a detailed analysis of  XMM$-$Newton spectra of OJ~287 spreading 15 yr. KOM21 also presented their  achieved  findings  from $Swift$ UVOT and XRT observation of OJ~287, which begun  in 2015, along with all  public $Swift$ information after 2005. During this period, OJ~287 was found in an {\it ``extreme''} low-hard  state (LHS) and outburst high-soft states (HSS). In addition, they established that the OJ~287  X-ray spectra were  highly variable and passed all states seen in blazars from  a {\it ``flat''} LHS through an intermediate state (IS) to exceptionally a soft  steep (ST) state.  KOM21 found that these  spectra can be made up using  the following parts: Inverse Compton (IC) radiation which is prevailing in the LHS, very soft  radiation that turned into  extremely powerful when  OJ~287 became  more luminous.  KOM21 claimed that  their 2018 XMM$-Newton$  measurements, almost contemporary with the EHT examination of OJ~287, were  well characterized  by a model with a hard IC component with the photon index  $\Gamma \sim 1.5$ and a soft  component.  It is important to emphasize  that they concluded that  that the  LHS spectra limited any long-lived accretion disc/corona contribution in X-rays  and related  to a very low value of $L_x/L_{Edd} < 5.6 \times 10^{-4}$ (for $M_{BH}$ of the  primary  $\sim 1.8\times 10^{10} M_{\odot}$). 
 
%%%%%%%%%%%%%%%%%%%%%%%%%%%%%%%%%%%%%%%%%%%%%%%
%
% FIGURE 2
%
%%%%%%%%%%%%%%%%%%%%%%%%%%%%%%%%%%%%%%%%%%%%%%%%%%%%
\begin{figure*}
\centering
\includegraphics[scale=0.70,angle=0]{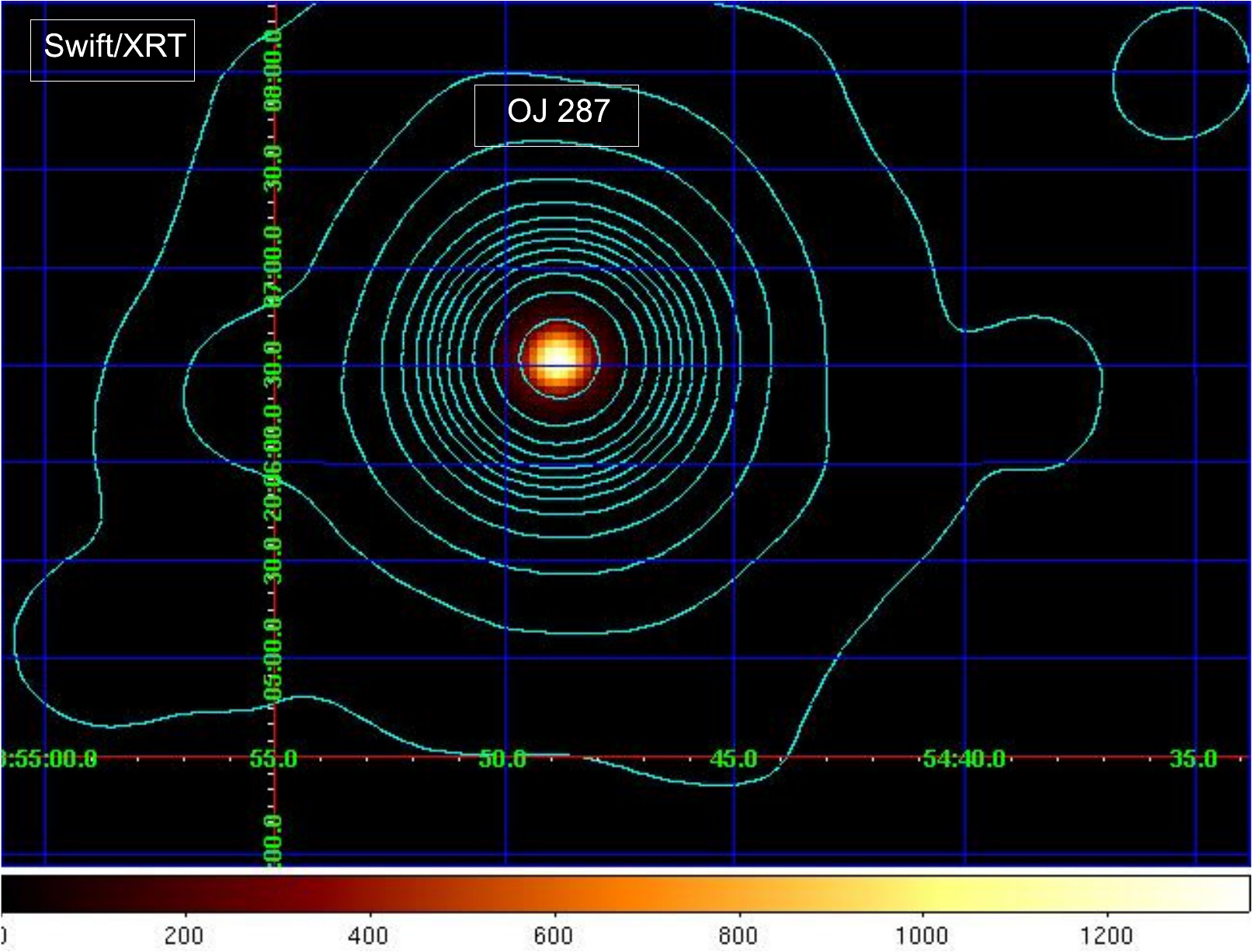}
\caption{A $Swift$ X-ray image on May 20, 2005 -- June 13, 2018 with 455 ks exposure.  
Contours correspond to fourteen logarithmic intervals with respect to the brightest pixel.
}
\label{imagea}
\end{figure*}

%%%%%%%%%%%%%%%%%%%%%%%%%%%%%%%%%%%%%%%%%%%%%%%
%
% FIGURE 3
%
%%%%%%%%%%%%%%%%%%%%%%%%%%%%%%%%%%%%%%%%%%%%%%%%%%%%
 
\begin{figure*}
\centering
\includegraphics[scale=0.87,angle=0]{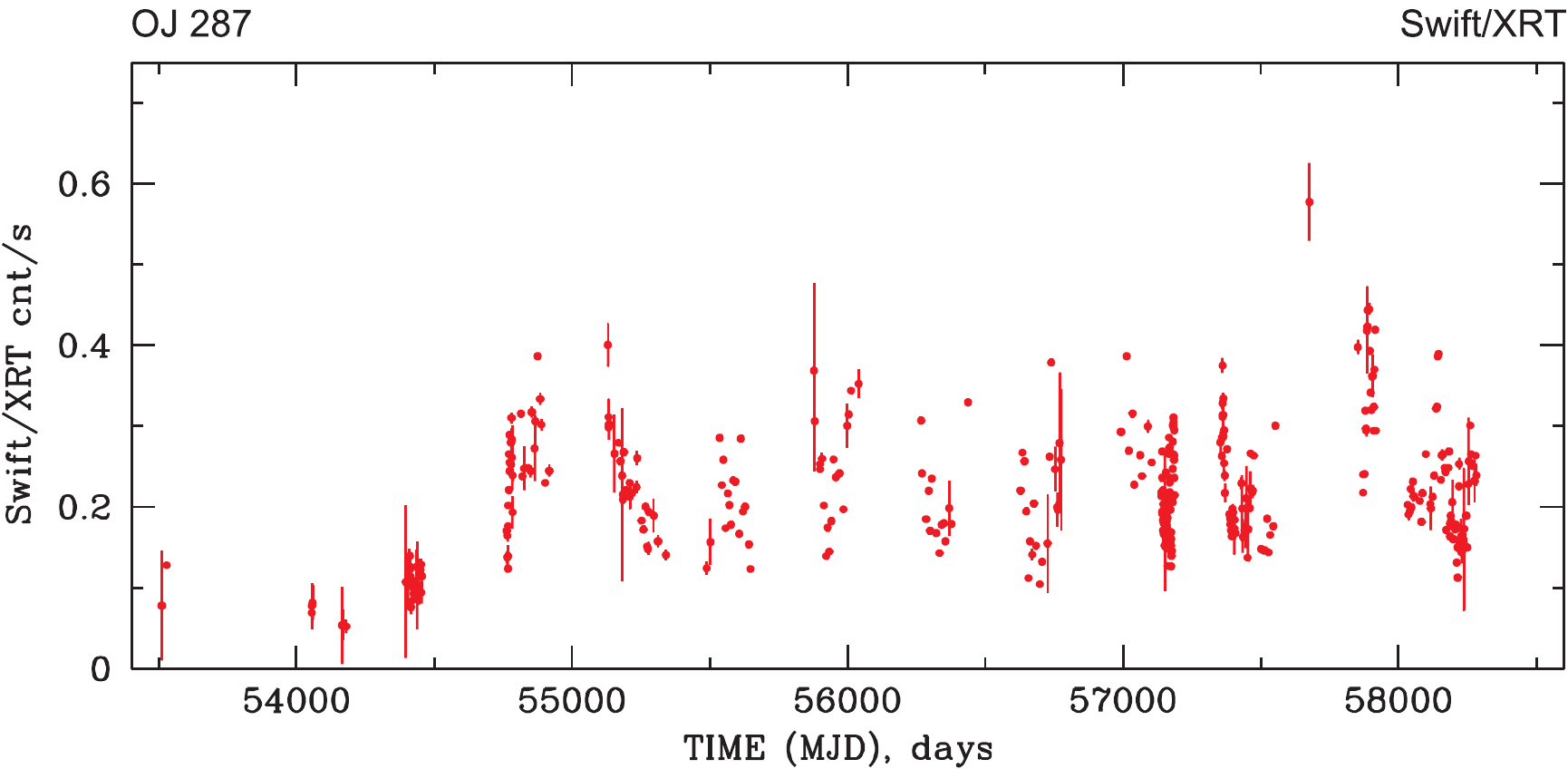}
\caption{Evolution of XRT/$Swift$ count rate during the 2005 -- 2018 observations of OJ~287.
}
\label{Swift_lc}
\end{figure*}

%%%%%%%%%%%%%%%%%%%%%%%%%%%%%%%%%%%%%%%%%%%%% 
%
%  FIGURE 4 spectra
%
%%%%%%%%%%%%%%%%%%%%%%%%%%%%%%%%%%%%%%%%%%%%%          

 \begin{figure*}
 \centering
\includegraphics[width=18cm]{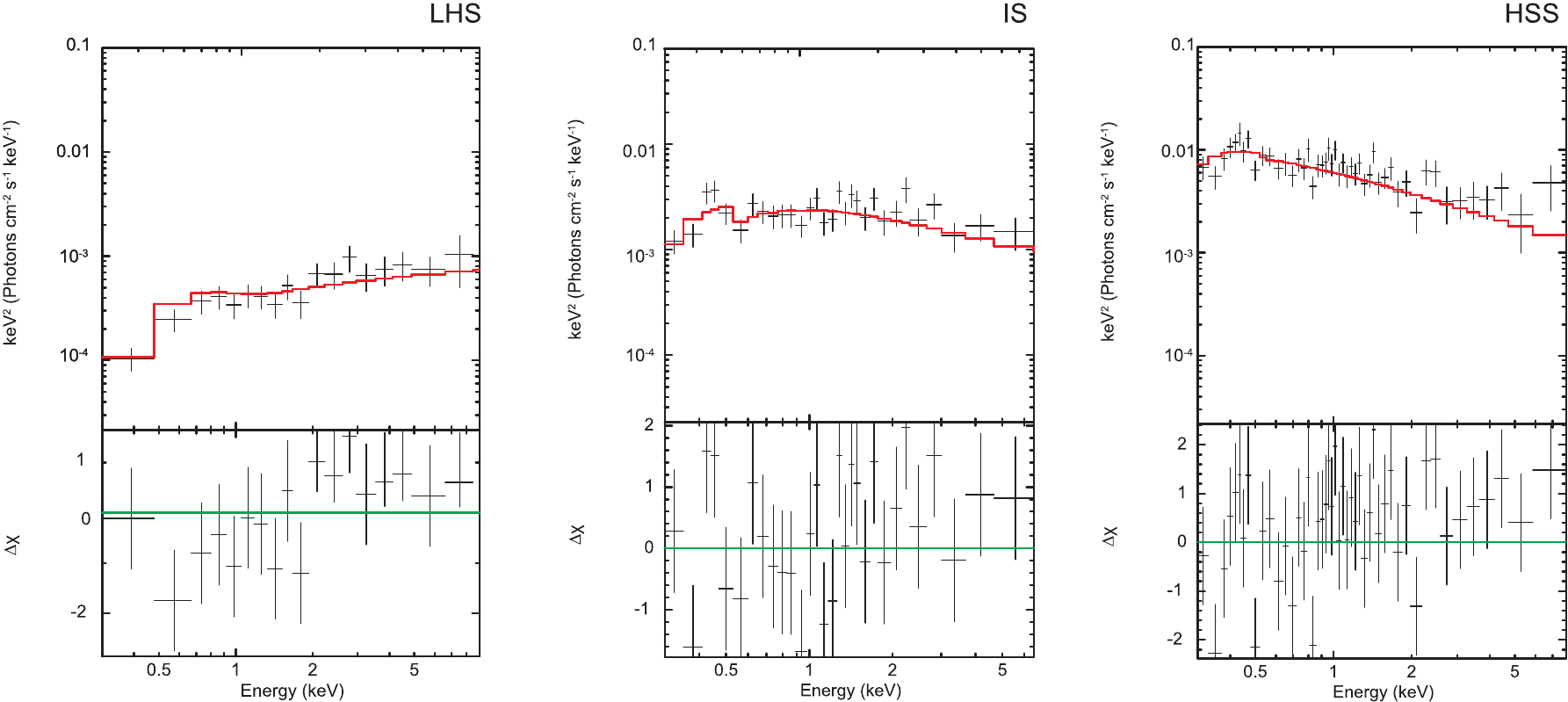}
   \caption{Three representative spectra of OJ~287 from {\it Swift} data with best-fit modelling for the LHS (ID=00035011001), IS (ID=00088085001)  and HSS (ID=00034934051)  states in units $E*F(E)$ using  {\tt tbabs*bmc} model.  The data are denoted by $black$ {crosses}, while the spectral model presented    by $red$ histogram. 
}
\label{3_spectra}
\end{figure*}

 We are suggesting a revision of this  claim by KOM21 reanalyzing  data of X-ray telescope (XRT) onboard of $Swift$ observatory  for OJ~287 and applying  the ST09 scaling method  to these data  in order to make a BH  mass estimate. Using this  method requires an accurate knowledge of the distance to the source OJ~287.  The method was proposed back in 2007 by \cite{st07}, hereafter ST07 and by ST09. It is worth noting that  there are two   scaling methods:  based on the correlation between the photon index, $\Gamma$ and  the quasi-periodic oscillation  (QPO) frequency,   $\nu_L$; and one  based on the correlation between $\Gamma$ and  normalization of the spectrum proportional to $\dot M$. If for the first {method} ($\Gamma-\nu_L$) to estimate a BH mass  the  the distance to the source is not required (ST07), while for the second {method},  $\Gamma-\dot M$  (see ST09) the source distance and  the { inclination}  of the accretion disk relative to the Earth observer  are needed.

%%%%%%%%%%%%%%%%%%%%%%%%%%%%%%%%%%%%%%%%%%%%%%%%%%%%
%
%   TABLE   1  Swift data
%
%%%%%%%%%%%%%%%%%%%%%%%%%%%%%%%%%%%%%%%%%%%%%%%%%%%%

\begin{table*}
 \caption{The list of $Swift$ observations of OJ~287 used in our analysis}              % title of Table
 \label{tab:list_Swift}      % is used to refer this table in the text
% \centering                                      % used for centering table
 \begin{tabular}{l l l l l c}          % centered columns (4 columns)
 \hline\hline                        % inserts double horizontal lines
  Obs. ID& Start time (UT)  && End time (UT) &MJD interval \\    % table heading
 \hline                                   % inserts single horizontal line
00020117001   & 2009 October 28 17:05:58   && 2009 October 29    & 55132.7 -- 55132.8 &\\
00030901(001-002,004-015,017-026,029-039, & 2007 March 6          && 2016 Oct 15    & 54165 -- 57676    & \\
041-048,050-051,053-061,063-064,066-067,&           &&     &     & \\
069,071-089,091,093-099,101-109,111-117,&           &&     &     & \\
119-122,124-128,130-146,148-150,152-155,&           &&     &     & \\
157-162,164-165,167-178,181-196,198-215,&           &&     &     & \\
217,219-220)&           &&     &     & \\
00035011(001,003) & 2005 May 20 && 2005 June 7& 53510 -- 53528 & \\
00035905(001-003)   & 2006 Nov 16           && 2006 Nov 18    & 54055.6 -- 54026 & \\
00090015(002-012) & 2008 Oct 24 && 2009 Sep 16 & 54763 -- 55090 &\\
00090086(001-003) & 2010 Jan 13   && 2010 Mar 23 &   55209 -- 55728 &\\  
00033756(002-020,022-068,073,075-086,088-089)  & 2015 Apr 28           && 2015 Jun 14          & 57140 -- 57187 &\\
00034934(051-053,055-070,072-080,082-083,& 2017 Feb 15           && 2018 Jun 13          & 57799 -- 58282 &\\
086-089,091,093-096,098-113,115-123,125  &       &&           &  &\\
127-143,145-147)  &       &&           &  &\\
00088085(001) & 2017 Apr 9 16:18:13  && 18:06:53& 57852 -- 57852.7 &\\
00088667001 & 2018 Apr 20 15:35:14  && 17:15:53& 58228 -- 58228.7 &\\
 \hline                                             %inserts single line
 \end{tabular}
%\\ $^{(a)}$ \cite{kom21}.
 \end{table*}

%%%%%%%%%%%%%%%%%%%%%%%%%%%%%%%%%%%%%%%%%%%%%%%%%%%%%%%%%%%%%%%
%
% TABLE 2 - BH MASS DETERMINATION
%
%%%%%%%%%%%%%%%%%%%%%%%%%%%%%%%%%%%%%%%%%%%%%%%%%%%%%%%%%%%%%%%
\begin{table*}
 \caption{BH masses and distances}
 \label{tab:par_scal}
 \centering 
 \begin{tabular}{lllllll}
 %\begin{tabular}{llllllll}
 \hline\hline                        % inserts double horizontal lines
Reference sources   & $m_r$ (M$_{\odot})$ & $i_r^{(a)}$ (deg) & $N_r$ ($L_{39}/d^2_{10}$) & $d_r$ (kpc) \\
      \hline\hline
XTE~J1550--564$^{(1)}$ &   10.7$\pm1.5$ &  72  & 1. &  3.3 $\pm0.5$ \\       %\cite{st09}  ST09
H ~1743--322$^{(2)}$  &  13.3$\pm3.2$ & 70 & 0.19 &  $9.1\pm1.5$\\   
4U~1630--47$^{(3)}$ & 10$\pm 0.1$& 70& 0.12 & $10\pm 1$\\
ESO 243-49$^{(4)}$ & $(7.2\pm 0.7)\times10^4$ & 75 & $4.2\times 10^{-6}$ & $ (95\pm 10)\times10^3$\\
M101 ULX-1$^{(5)}$ & $(3.7 \pm 0.6)\times10^4$ & $18$ & $3\times 10^{-4}$ & $(6.9\pm 0.7)\times10^3$\\
GRS 1915+105 $^{(6)}$ & $12.4\pm 2$ & $70$ & $0.2$ & $8.6\pm 2 $\\
 \hline\hline                        % inserts double horizontal lines
Target source   & $m_{t}$ (M$_{\odot}$) & $i_t^{(a)}$ (deg) & $d_t^{(b)}$ (kpc) &    \\
      \hline
OJ 287 & $\sim1.25\times(1\pm 0.45)\times10^8$ &   50  &  1.037 Gpc     &   that  using  XTE~J1550-564- as a reference source \\
 OJ 287  & $\sim1.25\times(1\pm 0.45)\times10^8$ &  50 &  1.037 Gpc    &    that  using  H ~1743--322  as a reference source \\ 
OJ 287  & $\sim1.25\times(1\pm 0.45)\times10^8$ &  50 &  1.037 Gpc    &    that  using  4U~1630--47  as a reference source\\
OJ 287  & $\sim1.25\times(1\pm 0.45)\times10^8$ &  50 &  1.037 Gpc    &    that  using  ESO 243-49  as a reference source\\
OJ 287  & $\sim1.25\times(1\pm 0.45)\times10^8$ &  50 &  1.037 Gpc    &    that  using  M101-ULX-1   as a reference source\\
OJ 287  & $\sim1.25\times(1\pm 0.45)\times10^8$ &  50 &  1.037 Gpc    &    that  using  GRS 1915+105   as a reference source\\
OJ 287  & Final  estimate                                      &  50 &  1.037 Gpc    &   as a standard deviation for a mean: \\% $0.45/6^{1/2}=0.18$\\
           &  $\sim1.25\times(1\pm 0.18)\times10^8$ &      &      &   $0.45/6^{1/2}=0.18$\\
\hline
%\\%, $\ge$ 15$^e$\\%$\sim$ 7--21 \\
% \hline                                             %inserts single line
 \end{tabular}
\\
(a)  System inclination in the literature and  
(b) source distance found in literature. %(c) scaling value found by ST09; 
%(c) scaled value found by $\Gamma-\nu_{L}$ correlation and (d) scaled value found by $\Gamma-N_{bmc}$ correlation of the present paper.
(1) ST09;  \cite{Orosz2002,SanchezFernandez99,Sobczak99};
(2) ST09;
(3) \cite{STS14}; % Seifina et al. (2014);
(4) \cite{tsei16b}; %  Titarchuk \& Seifina (2016a);
(5)  \cite{TS16} and  %Titarchuk \& Seifina (2016b);
(6) \cite{ts09}. %Titarchuk \& Seifina (2009).
\\
 \end{table*}
 
 For both methods, it is necessary that  the source shows a change in spectral states during the outburst  and a characteristic behavior of $\Gamma$. A monotonic increase of $\Gamma$   with $\nu_L$ {or} $\dot M$ in the  LHS$\to$IS$\to$HSS transition and reaching a constant level  ({saturating}) at high values of  $\nu_L$ or $\dot M$. Then $\Gamma$ monotonic decreases during HSS$\to$IS$\to$LHS transition when the outburst decay. The saturation of $\Gamma$ (so-called ``$\Gamma$-saturation phase'') during  outburst is a specific signature  that this particular object contains a BH~\citep{tz98}. Indeed,  the $\Gamma$-saturation phase can be caused only by an accretion flow converging to the event horizon of a BH (see the Monte-Carlo simulation results in  \cite{LT99,LT11}). Then, it makes sense to compare BH sources that have the same $\Gamma$-saturation levels.  In the second {method} ($\Gamma-\dot M$), it is assumed that  BH luminosity is directly proportional to $\dot M$ (and, consequently, to the mass of the central BH), and inversely proportional to the squared distance to the source. Thus, { we can determine the BH mass by comparing the corresponding tracks   
  $\Gamma-\dot M$ for} a pair of sources with BHs, in which all parameters are known for one  source, and for the other  source parameters except  a BH mass are known (for more details on the scaling method, see \cite{tss10,ST10} and ST09). 

The scaling method has a number of advantages in determination of a BH mass, compared to other methods.   Calculation of the X-ray spectrum originating in the innermost part of the source  based on first-principle (fundamental) physical models, taking into account the Comptonization of the soft disk photons by hot electrons of internal disk part and in a converging flow to a BH. In fact, in the case of a $10^8 M_{\odot}$ BH  the disk peak temperature is relatively low, $ kT_s<1~{\rm keV}/(M_{BH}/10M_{\odot})^{1/4}$, i.e. about 20--100 eV and its thermal peak is in the UV energy range [see details of our observation  in \S 3 and  \cite{ss73}]. It is worth noting that using this scaling method for a BH mass estimate \cite{TS16,STV17,TS17,STU18,SCT18,TSSO20} % \cite{STV17}, \cite{STU18}, \cite{SCT18}, and \cite{TSSO20} 
applied it   to intermediate mass BHs  and SMBHs.   

In this paper, based on {\it Swift} data analysis, we estimate a BH mass in 
OJ~287 using the scaling technique. In \S 2 we provide details of our data analysis, while in \S  3 
%\ref{sp_analysis} 
we present a description of the spectral models used for fitting these data. In \S 4 we focus on  the interpretation of our observations. In \S 5 we focused on the construction  of the power density spectra (PDS) and its interpretation. In  \S  6 we  discuss  the  main results of the paper. In \S 7  %\ref{conclusions} 
we present our final conclusions.

\section{ Data  Reduction  \label{data}}
\label{data}

{ Using {\it Swift}/XRT data in 0.3--10 keV energy range} we studied a total of 385 observations of OJ~287 during its flaring events from 2005 to 2018.  The data used in this paper are public and available through the GSFC public archive at https://heasarc.gsfc.nasa.gov. In Table~\ref{tab:list_Swift}  we report the log of observations for OJ~287 used in our study.  We must admit that not all of the flare events may related to the secondary BH -- disk  interactions.  Thus, it may be challenging to disentangle them from other flaring that goes on in OJ 287 all the time, completely unrelated to any binary's presence.

Data were processed using the HEASOFT v6.14, the tool {\tt xrtpipeline} v0.12.84 and the calibration files (CALDB version 4.1). The ancillary response files were created using {\tt xrtmkarf} v0.6.0 and exposure maps generated by {\tt xrtexpomap} v0.2.7. Source events were accumulated within a circular region with radius of 47{\tt''} centered at the position of OJ~287 [$\alpha=08^{h}58^{m}47^s.15$ and $\delta=+20^{\circ} 08{\tt '} 00{\tt ''}.5$, J2000.0]. We used XRT data both in the Windowed Timing (WT) mode ($\ge$ 1 count/s) and in the Photon Counting %(PC) 
mode for the remaining observations when the X-ray source became sufficiently  faint. The background was estimated in a nearby source-free circular region of 118{\tt''} radius. 

Using  the {\tt xselect} v2.4 task, source and background light curves and spectra were generated. 
Spectra were re-binned with at least 10 counts in  in each energy bin using the {\tt grppha} task in order to apply $\chi^2$-statistics. We also used the online XRT data product generator\footnote{http://www.swift.ac.uk/user\_objects/} to obtain the image of the source field of view in order to make a visual inspection and to get rid of  a possible contamination from nearby sources  \citep{Evans07,Evans09}. 
{\it Swift}/XRT (0.3 -- 10 keV) image of OJ~287 field of view is presented in Fig.~\ref{imagea} and demonstrates absence of the X-ray jet-like (elongated) structure as well as the minimal contamination by other point sources and diffuse emission within a region of 120{\tt''} radius around OJ~287. We used $Swift$ observation of OJ~287 (2005 -- 2018) extracted from the HEASARC archives and found that these data  cover a wide range of X-ray luminosities.  

Before to proceeding with details of the spectral fitting we study a long-term behavior of OJ~287, 
in particular, its activity patterns. We present a long-term X-ray light curve of OJ~287 detected by the XRT onboard  of the {\it Swift} over 2005 -- 2018 (see Fig.~\ref{Swift_lc}). 

Note, that this X-ray light curve makes it rather difficult to judge the 12-yr periodicity found earlier from optical observations. But it can be unequivocally stated that the object OJ~287 has become active over the past 10 years and shows sporadic X-ray activity (e.g., MJD 54500--58200, see Fig.~\ref{Swift_lc}).

\section{Analysis and Results \label{results}}

In this section we present the results of spectral analysis during the  outburst of  OJ~287 observed by $Swift$/XRT. We  analyze  how the X-ray spectrum of the source behaves, in particular, $\Gamma$ during 
the outburst.

In our paper, we adhere to the scenario in which OJ~287 is a binary system consisting of two black holes with a larger and smaller mass (Fig.~\ref{picture}). In this case, a heavier BH (primary BH) is surrounded by a powerful accretion disk, and a lighter BH (secondary BH) orbits around the primary BH in a plane different from the equatorial plane of the primary BH accretion disk, crossing it twice during the orbital period (12 years). We also assume that when the secondary BH passes through the disk around the primary BH, ``tidal disruption'' of nearby parts of the disk occurs, due to which the partial disruption and subsequent matter replenishment of the accretion disk around the secondary BH is possible [see a similar process in \cite{Chan21}]. In this case, a powerful transient disk develops around the secondary SBH with subsequent accretion of the material of the transient disk onto the secondary BH. This provides an increase in luminosity in the form of an ouburst in the X-ray/optical/radio bands.

\subsection{Spectral analysis \label{sp_analysis}}

To fit the energy spectra of this source we used an {\tt XSPEC} model consisting of  the Comptonization (bulk-motion Comptonization, hereafter  BMC) component [see \cite{tz98, LT99}]. We also use a multiplicative {\tt tbabs} model \citep{W00} which takes into account  absorption by neutral material. We assume that  accretion onto a BH is described by two main zones [see, for example, Fig.~1 in \cite{TS21}]: a geometrically thin accretion disk (e.g. the standard Shakura-Sunyaev disk, see SS73 and a transition layer (TL), which is an intermediate link between the accretion disk, and a converging (bulk) region (see \cite{tf04}), that is assumed  to exist, at least, below 3 Schwarzschild radii, $3R_S = 6GM_{\rm BH}/c^2$. The spectral model parameters are the equivalent hydrogen absorption column density $N_H$; the photon index $\Gamma$;  $\log (A)$ is related to the Comptonized factor $f$ [$={A}/{(1+A)}$]; the  color temperature and normalization of the seed photon blackbody component, $kT_s$ and   $N$, respectively.  The parameter $\log(A)$ of the BMC component is fixed at 2 when the best-fit $\log(A)\gg 1$. In fact, for a sufficiently high $\log(A)\gg1$ (and, therefore, a high value A), the illumination factor $f = A/(1 + A)$ becomes a constant value close to 1 (that is, the same as in the case of $\log(A) = 2$). $N_H$ was fixed at the Galactic absorption level of {$2.49\times 10^{20}$ cm$^{-2}$~\citep{W00}. 

%%%%%%%%%%%%%%%%%%%% 
%
%  FIgure 4- GAMMA - NORM
%
%%%%%%%%%%%%%%%%%%%% 

 \begin{figure*}
 \centering
\includegraphics[width=10cm]{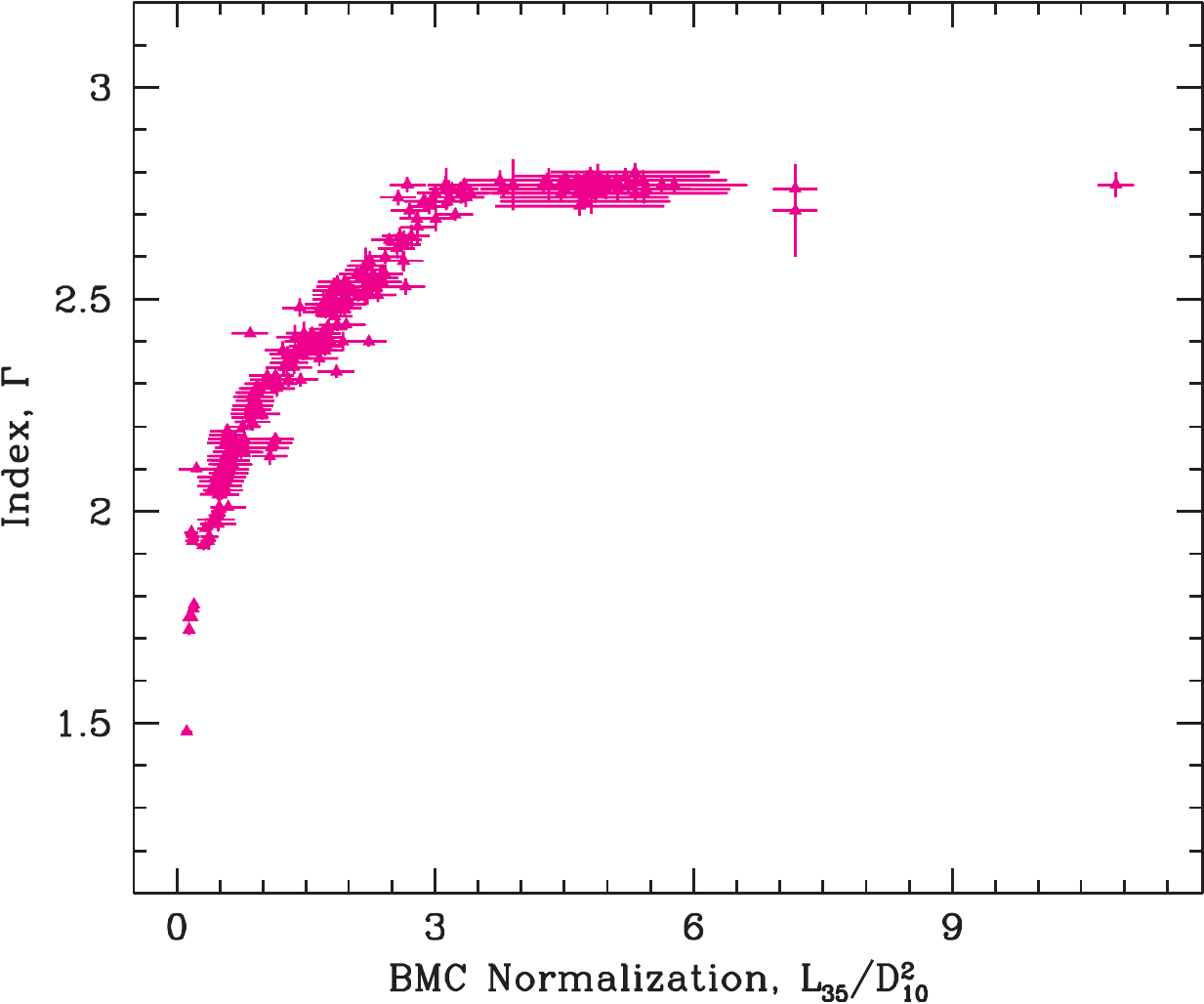}
   \caption{
Correlation of the photon index $\Gamma$ ($=\alpha+1$) versus  the BMC normalization, $N$ (proportional to mass accretion rate) in units of $L_{39}/D^2_{10}$. 
}
\label{saturation}
\end{figure*}

Similarly to the  bbody XSPEC model, the  normalization  is a ratio of the source (disk) luminosity $L$ to the square of the distance $d$  (ST09, see  Eq.~1 there): 

\begin{equation}
N=\biggl(\frac{L}{10^{39}\mathrm{erg/s}}\biggr)\biggl(\frac{10\,\mathrm{kpc}}{d}\biggr)^2.
\label{bmc_norm}
\end{equation}  

This encompasses  an important property of our model. Namely, using this model one can correctly
evaluate  normalization of the original ``seed'' component, which is presumably a correct $\dot M$ indicator~\citep{ST11}. In its turn 

\begin{equation}
L = \frac{GM_{BH}\dot M}{R_{*}}=\eta(r_{*})\dot m L_{Ed}.
\label{bmc_norm_lum}
\end{equation}  
Here $R_{*} = r_{*} R_S$ is an effective radius where the main energy release takes place in the disk, $R_S = 2GM/c^2$ is the Schwarzschild radius, $\eta = 1/(2r_{*})$, $\dot m = \dot M/\dot M_{crit}$ is the dimensionless $\dot M$ in units of the critical mass accretion rate $\dot M_{crit} = L_{Ed}/c^2$, and $L_{Ed}$ is the Eddington luminosity. For the formulation of the  Comptonization problem  one can see \cite{tmk97,tz98,LT99,Borozdin99,st09}.  

%%%%%%%%%%%%%%%%%%%%%%%%%%%%%%%%%
% 
%  FIgure 5 - SCALING 1550
%
%%%%%%%%%%%%%%%%%%%%%%%%%%%%%%%%%
\begin{figure*}
\centering
\includegraphics[width=12cm]{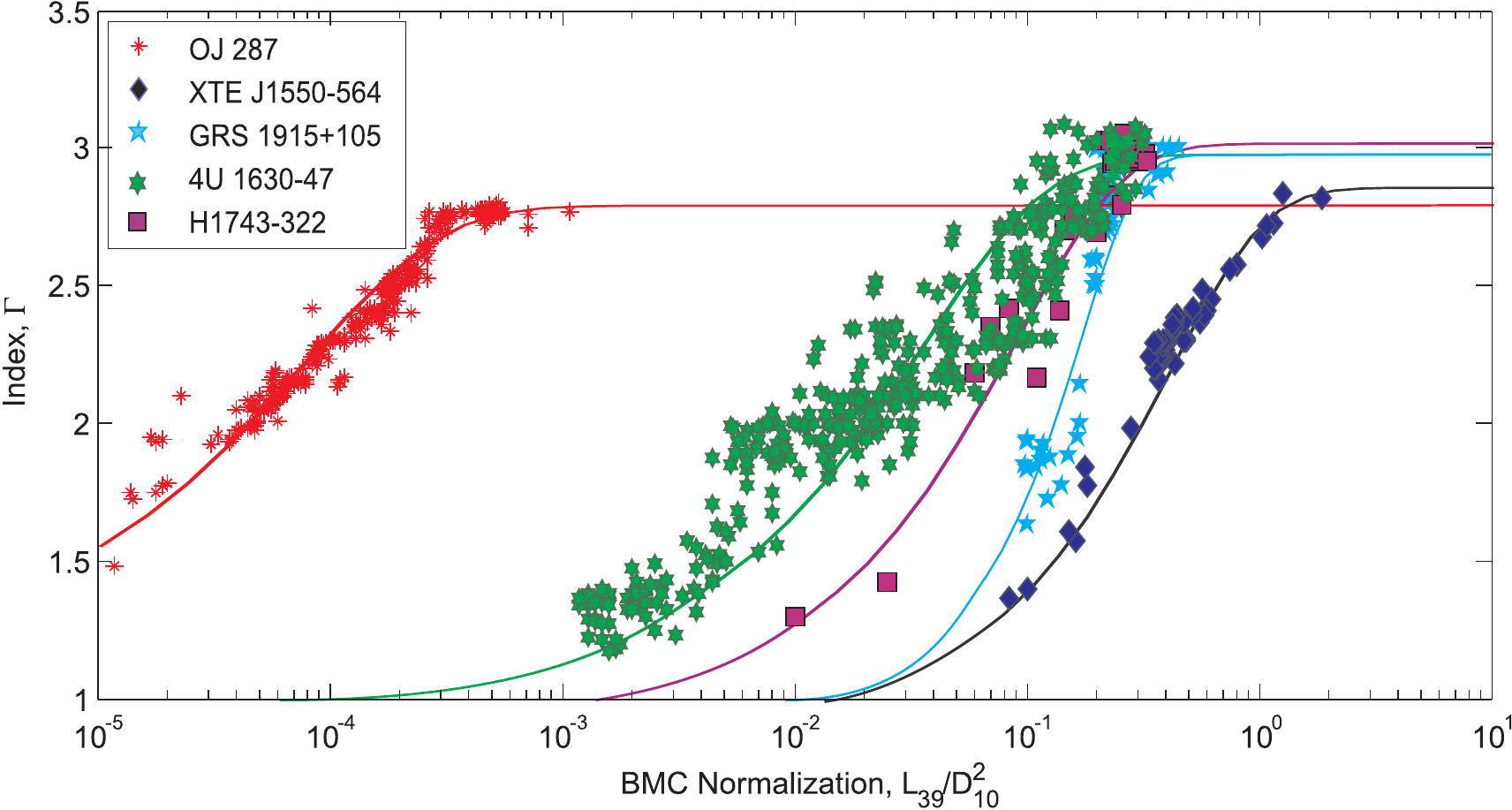} 

  \caption{Scaling of the photon index, $\Gamma$ versus the normalization $N_{BMC}$ for OJ~287 (red  points -- target source) using the correlation for the Galactic reference sources, XTE~J1550--564 (blue diamonds), H~1743--322 (pink squares), 4U~1630--47 (greed stars) and GRS~1915+105 (brigh blue stars). 
} 
\label{three_scal}
\end{figure*}

Spectral analysis of the $Swift$/XRT data { fits} of OJ~287, in principle, provides a general picture of the source evolution.  We can trace the change in the spectrum shape during the LHS--IS--HSS transition. 
In  Figure ~\ref{3_spectra}, we  show  three representative $E*F_E$ spectral diagrams %(red lines) 
for different states of OJ~287. 

We put together spectra of the LHS, IS, and HSS, to demonstrate the source spectral evolution from the low-hard to   high-soft to  states based on the $Swift$ observations. Data are presented here: at the left panel for LHS (taken from observation 00088085001), at the central panel for IS (00035011001) and at the right panel  for the HSS (taken from observation 00034934051) in units $E*F(E)$ fitted using  {\tt tbabs*bmc} model.  Periods of exposure  are  3.7, 1.9  and 1.1 ks,  respectively. 

The best-fit parameters in the HSS state (right panel) are $\Gamma$ = 2.77$\pm$0.02, $kT_s$ = 50$\pm$4 eV, $N$ = 7.2$\pm$0.6 L$_{35}$/d$^2_{10}$ and $\log(A)$ = 0.28$\pm$0.06  [$\chi^2_{red}$=0.98 for 265 degrees of freedom (dof)], while the best-fit model parameters for the IS state (central panel) are $\Gamma$ = 2.6$\pm$0.2, $kT_s$ = 40$\pm$2 eV, $N$ = 4.3$\pm$0.9 L$_{35}$/d$^2_{10}$ and $\log(A)$ = -0.72$\pm$0.08  [$\chi^2_{red}$=0.93 for 215 dof]; and, finally, the
best-fit model parameters for the LHS state (left panel) are $\Gamma$ = 1.7$\pm$0.3, $kT_s$ = 120$\pm$6 eV, $N$ = 0.19$\pm$0.05 L$_{35}$/d$^2_{10}$ and $\log(A)$ = --0.24$\pm$0.08 [$\chi^2_{red}$=1.06 for 237 dof]. A systematic uncertainty of 1\% is intended to represent the instrumental flux calibration uncertainty and  has been applied to all analyzed $Swift$ spectra. 

Analysis of  the $Swift$/XRT data fits (see  Figure~\ref{saturation}) showed that  $\Gamma$  monotonically increases from 1.5 to 2.8, when the normalization of the spectral  component (or $\dot M$) increases by a factor about  5. 

%%%%%%%%%%%%%%%%%%%%%%%%%%%%%%%%%%%
% 
%  FIgure 6- SCALING
%
%%%%%%%%%%%%%%%%%%%%%%%%%%%%%%%%%%%
 \begin{figure*}
 \centering
 \includegraphics[width=12cm]{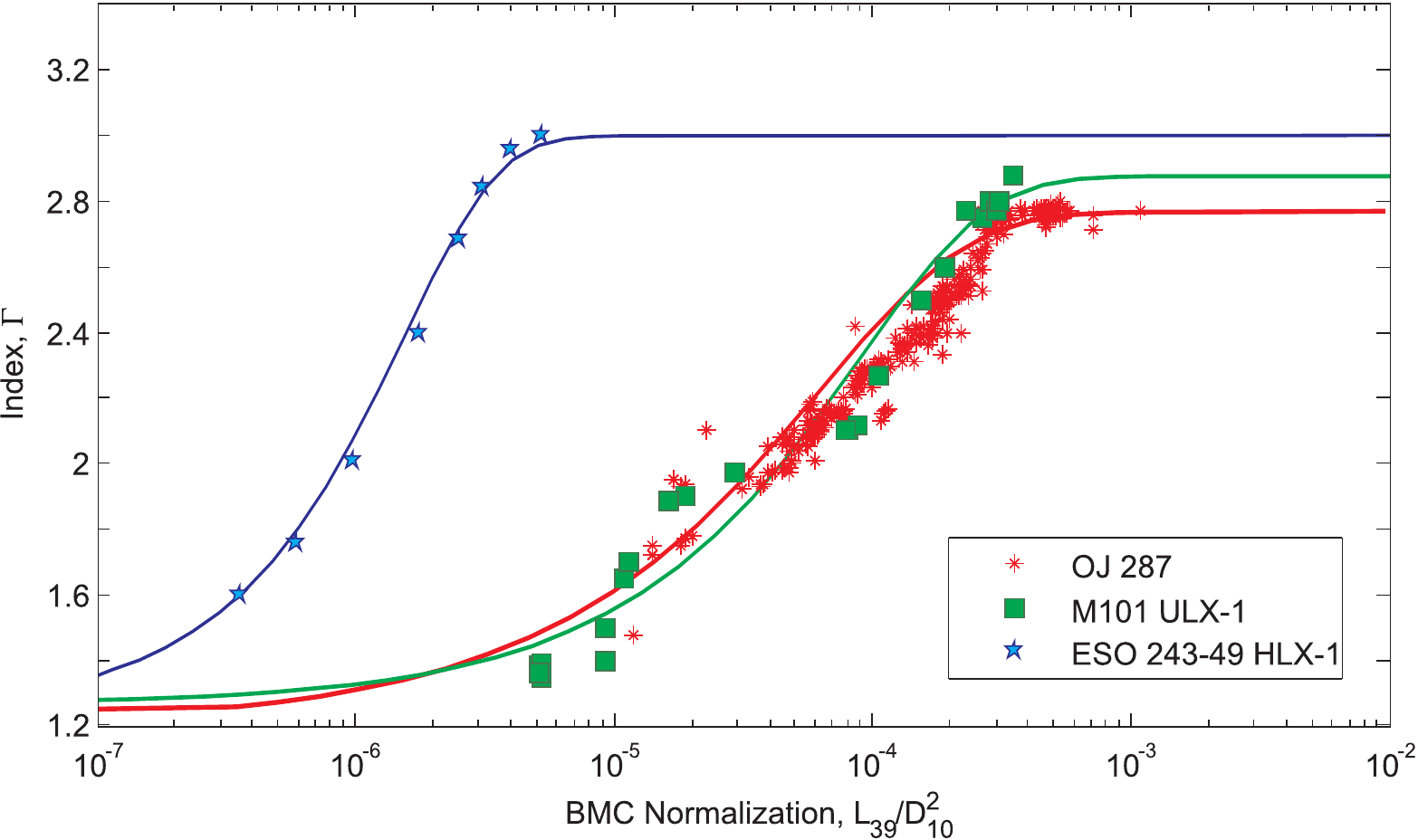}
      \caption{
Scaling of the photon index $\Gamma$ versus the normalization $N$ for OJ~287  (red line -- target source) using $\Gamma - N$ correlations for extragalactic sources, ESO~243--49 HLX--1 and M101~ULX--1 (blue  and green points).  
}
\label{three_scal_1}
\end{figure*}

%*******************
%              FIGURE 7
%*******************
 \begin{figure*}
 \centering
\includegraphics[width=15cm]{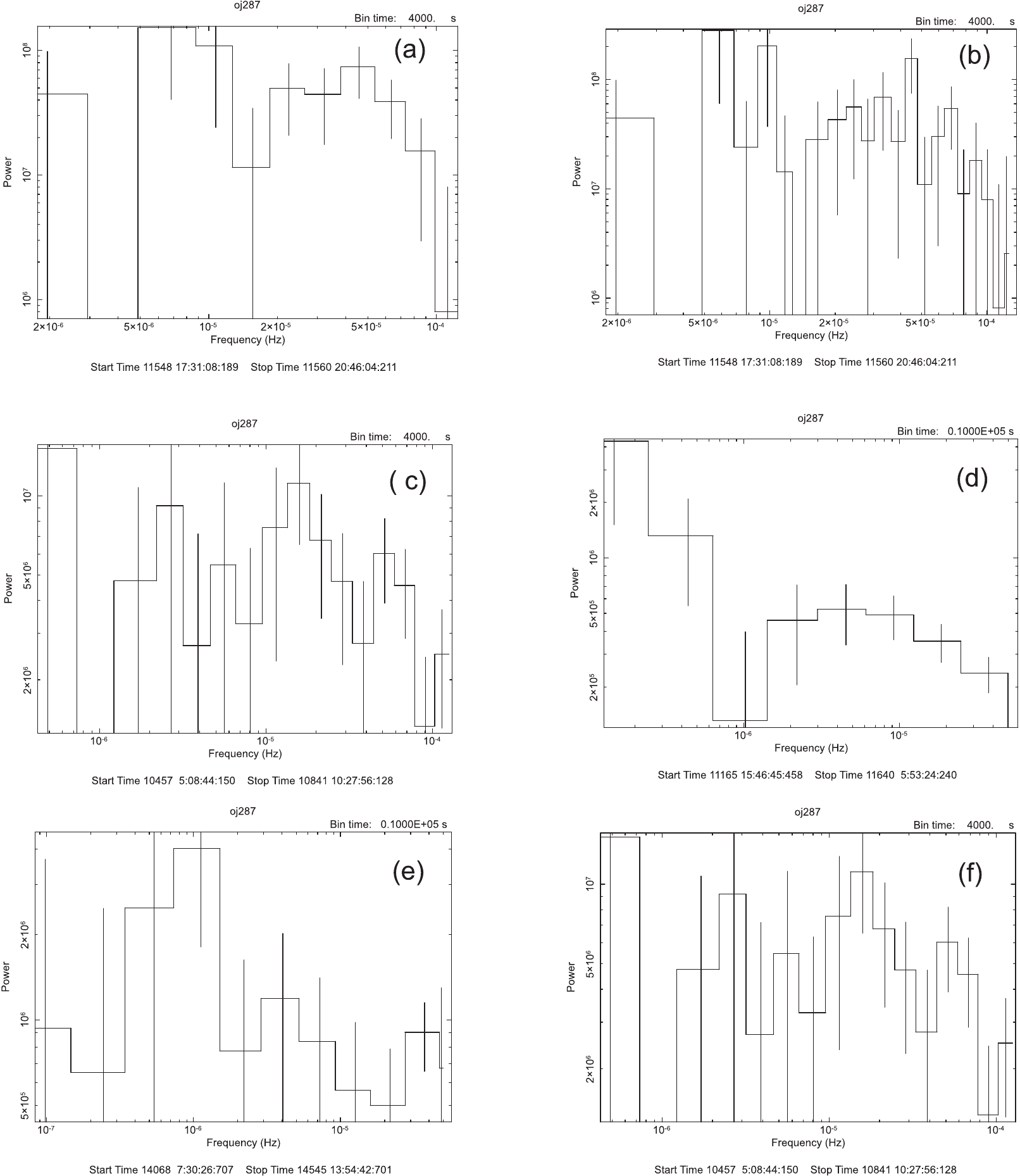}
\caption{ Evolution of  the power density spectrum %(PDS)  
of OJ~287 with time.  It is clearly seen the PDS evolution with the characteristic peaks in the range 
of  $10^{-6}$ and $10^{-5}$ Hz.
}
\label{Fig_PDS}
\end{figure*}

\section{A BH mass estimate of the OJ~287 secondary}
\label{mass_estimate}

Now, we use the scaling method for our estimate of the  BH mass,  $M_{oj287}$. Beforehand we do not know  what kind of a BH mass can be estimated using our  data for   OJ~287. The BH mass using the scaling method applies   the $\Gamma-N$ correlation (see  ST09 for details).  This method comes to ({\it i}) identifying 
for a pair of BHs for which the $\Gamma$ correlates with {increasing} normalization $N$ (which is proportional to mass accretion rate $\dot M$ and a BH mass $M$, see ST09, Eq.~(7)) and the saturation level, $\Gamma_{sat}$ are the same and ({\it ii}) calculating the scaling coefficient $s_{N}$ which allows  us to determine  a BH  mass of  the target object. We also should  emphasize that one needs  a ratio of  distances  for the target and reference sources in order to estimate  a BH mass using  the following equation for the  scaling coefficient 
\begin{equation}
 s_N=\frac{N_r}{N_t} =  \frac{m_r}{m_t} \frac{d_t^2}{d_r^2}{f_G}
\label{mass}
\end{equation}
where $N_r$, $N_t$ are normalizations of the spectra, $m_t=M_t/M_{\odot}$,  $m_r=M_r/M_{\odot}$ are the dimensionless  BH masses with respect to solar, $d_t$ and $d_r$  are distances to  the target and reference sources, correspondingly.  A   geometrical factor, $f_G=\cos i_r/\cos i_t$ where $ i_r$ and $ i_r$ are the disk inclinations for   the reference  and target sources, respectively [see ST09, Eq.~(7)].

We found that  XTE~1550--564, H~1743--322, 4U~1630--47, GRS~1915+105, ESO~243--49 and M101~ULX--1 can be used as the reference sources because these sources met all aforementioned  requirements to estimate a BH mass of the target source OJ~287 [see  items (i) and (ii) above]. 

In Figure \ref{three_scal}  we demonstrate how the photon index $\Gamma$ evolves with normalization $N$ (proportional to the mass accretion rate $\dot M$) in the Galactic source XTE~1550--564  and OJ~287  where  $N$  is presented in the units of $L_{39}/d^2_{10}$ ($L_{39}$ is the source luminosity in units of $10^{39}$ erg/s and $d_{10}$ is the distance to the source in units of 10 kpc).

As we show  in Fig. \ref{three_scal_1} that the correlations $\Gamma$ vs $N$ are self-similar for the target source (OJ~287) and two  M101 ULX--1 and ESO~243--49 HLX--1 are ultraluminous X-ray sources. Moreover, these three sources have almost the same index saturation level $\Gamma$ about 2.8. We estimated a BH mass for OJ~287  using the scaling method  (see e.g. ST09). In Figure~\ref{three_scal} we illustrate  how the scaling method works shifting one correlation vs another.     From these correlations we could estimate $N_t$, $N_r$ for OJ~287 and   for the reference sources (see Table~\ref{tab:par_scal}). A value of $N_t=2.4\times10^{-4}$,  $N_r$ in units of $L_{39}/d^2_{10}$  is determined in the beginning of the $\Gamma$-saturation  part [see Figs. \ref{three_scal} and \ref{three_scal_1},  ST07,  ST09, \cite{STS14,TS16,tsei16b,ts09}].
% Seifina et al. (2014); Titarchuk \& Seifina (2016a); Titarchuk \& Seifina (2016b) \& Titarchuk \& Seifina (2009)]. {STS14,TS16,tsei16b,ts09}

A value of   $f_G=\cos {i_r}/\cos{i_t} $  for the target and reference sources  can be obtained  using  inclination for OJ~287 $i_t=50^{o}$.  and for $i_r$   (see Table \ref{tab:par_scal}).
As  a result of the estimated target  mass (OJ~287), $m_t$ we find  that
\begin{equation}
m_t= f_G\frac{m_r}{s_N} \frac{d_t^2}{d_r^2} 
\label{mass_target1}
\end{equation}
where we use  values of $d_t=1.073$ Gpc . % (see Table \ref{tab:par_scal}).

 Applying   Eq.(\ref{mass_target1}), we can estimate  $m_t$ (see Table \ref{tab:par_scal})  and we  find  that the secondary BH mass  in OJ~287 is about $1.25\times(1\pm 0.18)\times10^8$ solar masses. To obtain this estimate  with appropriate error bars we need to consider  error bars for $m_r$ and $d_r$ assuming in the first approximation, errors for  $m_r$ and $d_r$ only.
 We rewrote Eq. (\ref{mass_target1}) as
 \begin{equation}
m_t(1+\Delta m_t/m_t)= f_G\frac{m_r}{s_N} \frac{d_t^2}{d_r^2}(1+\Delta m_r/m_r)(1+ 2 \Delta d_r/ d_r).
\label{mass_target_expand}
\end{equation}

Thus we obtained errors for the $m_t$ determination (see Table \ref{tab:par_scal}, second column for the target source) that 
\begin{equation}
\Delta m_t/m_t \sim \Delta m_r/m_r + 2 \Delta d_r/ d_r.
\label{mass_target_errors}
\end{equation} 

In order to calculate  the  dispersion $\cal D$ of the arithmetic  mean  $\bar{ m_t}$ for a BH  mass estimate using different reference sources $\cal D $ (see   Table \ref{tab:par_scal}), one should keep in mind  that 
\begin{equation}
{\cal {D}} (\bar{m_t})= D/n
\label{dispersion_mean}
\end{equation} 
where $D$ is the dispersion of $m_r$  using  each of the reference source and $n=6$ is a number of  the reference sources. As a result we obtained that  the mean deviation  of the arithmetic mean  
\begin{equation}
\sigma  (\bar{m_t})= \sigma/\sqrt{n}\sim 0.18
\label{sigma_arithm_mean}
\end{equation}
 and finally  we came to  that  (see also Table \ref{tab:par_scal})
 \begin{equation}
\bar{m_t} \sim1.25\times(1\pm 0.18)\times10^8 ~~~\rm {solar~masses}. 
\label{arithm_mean}
\end{equation}

It should be noted that in our calculations we assume the angle between the normal to the secondary disk
 and the line of sight to be about  50 degrees. However, this angle may be different. In fact, \cite{Dey21} and \cite{val21} argued that one should see the secondary disk almost face-on, namely, this angle $i_t$ is about zero. Consequently  the mass of the secondary should then be slightly lower, $\bar{m_t} \sim0.8\times10^8$ ~~~\rm {solar~masses}.

\section{Power density spectrum}
\label{PDS}
In Figure~\ref{Fig_PDS} we presented evolution of  the power density spectrum %(PDS)  
of OJ~287 with bintime of 4000 s %$3\times 10^5$s (top) and $0.1\times 10^5$s  (bottom)
 using {\it RXTE}/ASM data from 1997 to 2015 in 0.3--12 keV energy range.  We made  PDSs in the range of    $10^{-7}-10^{-4}$ Hz frequency  and subtracted the contribution because to Poissonian statistics. As it is seen from this Figure  the PDS undergoes  temporal evolution. If in  the upper plot we see a wide plateau  from $7\times10^{-6}$ to $10^{-5}$ Hz,  while in the lower panel it takes place in much wider of frequency range  from  $10^{-6}$ to $10^{-5}$ Hz. We can evaluate  the size of the Compton cloud (CC) emitting the emergent spectra  using the presented  plateaus.   The characteristic frequency  of  these plateaus  $\nu_{plat}$ can be estimated as
\begin{equation}
\nu_{plat}\sim V_{plas}/L_{CC}= 1.4\times10^8\frac{V_{plas}/(10^8 \rm {cm~s^{-1}})}{L_{CC}(\rm cm)} \rm{cm~s^{-1}},
\label{plat_frequency}
\end{equation} 
where $L_{CC}$ is the CC size and $V_{plas}=1.4\times10^8$ cm~s$^{-1}$ is a typical plasma (proton) velocity  in the CC  related to the plasma temperature  of order of 10 keV [see  e.g. \cite{st09}].
We use  a frequency $\nu_{plat}\sim10^{-5}$ Hz in order to estimate the CC size:
\begin{equation}
L_{CC}\sim 1.4 \times10^{13}\frac{V_{plas}/10^8 \rm {cm~s^{-1}}}{\nu_{plat}/10^{-5}~{\rm Hz}} {\rm cm}.
\label{CC_size}
\end{equation}
Using this CC size  $L_{CC}$ one can easily estimate   the appropriate  BH mass:
\begin{equation}
M_{oj287}\sim L_{CC}/R_{{\rm S}\odot} \sim 10^8 {\rm solar ~masses},  
\label{BH_estimate}
\end{equation}
which is, by order of magnitude,  close to our BH mass  value using the $\Gamma-$ mass accretion rate correlation [see  Eq. (\ref{arithm_mean}), 
Table \ref{tab:par_scal}} and  Fig. \ref{three_scal_1}]. 

\section{Discussion \label{discussion_OJ287}}
 In the previous chapter  using  the power spectra (see Figure \ref{Fig_PDS})  we estimated the CC size   as $L_{CC}\sim 1.4 \times10^{13}$ cm  where the X-ray emergent spectrum  was formed. Applying this  value  of $L_{CC}$ we confirmed a BH mass value of order $10^8$ solar masses  found using the  $\Gamma-$ mass accretion rate correlation (see Table \ref{tab:par_scal}, Figs. \ref{three_scal}-\ref{three_scal_1}, and section \ref{mass_estimate}).   This type of the  CC, with $L_{CC}$   of order of $10^{13}$ cm is definitely not related to a BH mass of order of  $10^{10}$ as KOM21 %\cite{kom21}, hereafter Kom21, 
claimed. 
 
KOM21  also noted  that their inferred X -ray luminosity with respect to the Eddington one  $L_x/L_{\rm Edd}<6\times10^{-4}$ was too small for $M_{\rm BH~primary}\sim 1.8\times 10^{10} M_\odot$.   Furthermore,  they  mentioned, in their section 4.3  a possibility that $L_{\rm x, iso}=1.3\times10^{45} $ erg s$^{-1}$ can be associated with the secondary of a BH mass  $M_{\rm BH,~secondary} \sim 1.5\times10^8M_{\odot}$. In KOM21  the authors presented a profound  spectral evolution from the low to high states (see their Figs.~ 4--7). They  correctly  claimed using their spectral results  that all spectral states observed in OJ 287  in the 0.3--10 keV  band evolved  from rather flat to ultra-steep    $(\Gamma_x=1.5-2.8)$. We  confirmed  this kind of  spectral behavior  in our Figures \ref{3_spectra}-\ref{saturation} that  $\Gamma$  significantly evolved  depending on Normalization (proportional to the mass accretion rate)  from 1.5 to 2.8 too.

One can argue that we did not realize, that the emission from this BL Lac object, comes from a jet.  In fact,  we do not see any serious arguments for this statement. A particularly important point regards the fitting of $Swift$ X-ray spectra.  One can think that  our spectral  models contains  too many components and thus  cannot  fit to low-resolution $Swift$ data,  since there are then much more free parameters than actual independent data bins.  This is not the case because our continuum spectral model is  an XSPEC model consisting of the Comptonization  (BMC) component. The spectral model parameters are the equivalent hydrogen absorption column density $N_H$; the photon index, $\Gamma$; $\log(A)$ is related to the Comptonized factor $f$;  the color temperature and normalization of the seed photon blackbody component, $kT_s$ and $N$, respectively. One can claim that simple power-law models are appropriate for $Swift$ spectral fitting. However, it is important to emphasize that a power-law itself is not a physical model but if one considers any up-scattering  model  (or just a particle  acceleration)  in the energetic cloud, which a particle energy is much greater than that of photons, then a power-law  is formed   (see a proof of this statement in ST09).

In the light of the obtained results, we can take an updated look at the discrepancy between the BH mass in the nucleus of the M87 galaxy, obtained using different methods. Namely, the discrepancy in the estimates of the BH mass in M87 $(3.5-6.5)\times 10^9 M_{\odot}$ [from its EHT radio image~\citep{Akiyama19} and based on the  gas dynamic analysis~\citep{Walsh13,Akiyama19} and stellar dynamics of M87~\citep{Akiyama19} using long-term optical observations~\citep{Gebhardt11}] and $6.5\times 10^7 M_{\odot}$ (by the method BH mass scaling from X-ray data~\citep{TSSO20}. \cite{TSSO20} reduced the BH mass in M87 by a factor of 100 compared to standard methods~\citep{Gebhardt11,Walsh13,Akiyama19}, using the timing analysis of the X-ray variability in M87. This discrepancy is difficult to explain only due to different ranges of radiation energy observations. Assuming that M87 contains a binary BH at its center \citep{Emami+Loeb20,Davelaar+Haiman22,Dou22}, an estimate of the BH mass by analyzing the power spectrum of M87 using a characteristic variability time of $5\times 10^{-7}$ s yields an estimate of the CC size in M87 of $L_{CC}\sim 2\times 10^{13}$ cm and the BH mass of $(6.5\pm 0.5)\times 10^7 M_{\odot}$. At the same time, these two SMBHs can be distinguished only by the scale of variability. In terms of this approach, it is possible that the BH mass measurement in M87 by \cite{TSSO20} refers specifically to the BH of smaller mass (secondary), in contrast to the approach of the gas and stellar dynamics as well as the EHT image analysis methods for an estimate of the primary BH mass or the total (primary + secondary) BH mass. This can be considered as a possible observational indication of the presence of a BH binary in M87. 

Recently, Ning Jiang and his colleagues~\citep{Jiang22} pointed out a similar picture of BH duality in the active galaxy nuclear. Namely, they argued for the probable presence of a BH binary in the Seyfert galaxy  SDSSJ143016.05+230344.4. \cite{Dou22} indicated that in this binary system, consisting of two SMBHs of different masses, a decay of the orital period was observed, which confirms the initial hypothesis of the presence of an eccentric SMBH binary in the center of this galaxy.

\section{Conclusions}
\label{conclusions}
The multi-wavelength outburst activity  in OJ~287 with the  X-ray telescope onboard the $Swift$  made a lot of questions  whether the source  contains one or two  BHs. It is  very important  to reveal  the   characteristics of this binary. In the presented paper we demonstrated  that  the OJ~287 X-ray spectra underwent  the state transition from the LHS to the IS and then to  the HSS (see Fig.~\ref{3_spectra}).   We obtained  that energy spectra in all spectral states can be modeled using  a product  of the  wabs  and a BMC Comptonization component. 

Moreover, we discover in OJ~287  the correlation of the index $\Gamma$ with normalization, $N$ (proportional to the disk mass accretion rate $\dot M$, see Fig.~\ref{three_scal_1}) similar to those established in BH Galactic sources by ST09.  We found that $\Gamma$ increases monotonically with $\dot M$ from the LHS  to  IS and HSS, and then saturates at $\Gamma\sim$ 2.8.  This can be  considered as  observational evidence of the presence of a BH   in OJ~287. Based on this correlation, we apply the scaling method of ST09 to estimate a BH mass is about   $1.25\times 10^8$  solar masses, using the well-studied Galactic  X-ray BHs, XTE~1550--564, H~1743--322, 4U~1630--47,  GRS~1915+105 and extra-galactic BHs ESO~243--49 and M101~ULX--1  as  reference sources.  
 
Also using the power spectrum analysis we inferred the size of the Compton cloud  $L_{CC}\sim 10^{13}$ cm where X-ray spectra were formed. Using this value of   $L_{CC}$ we confirmed that a BH mass of the secondary in  OJ~287  was of order of $10^8$ solar masses consistent with  the index, $\Gamma-$correlation (the scaling method)  with respect to the mass accretion rate.

~~~~~~
\section*{Acknowledgements}
We acknowledge support from UK $Swift$ Science Data Centre at the University of Leicester for supplied data. 
We thank the anonymous referee for the careful reading of the manuscript and for providing valuable comments.  

\section*{Data availability}
This research has made using the  data and/or software provided by the High Energy Astrophysics Science Archive Research Center (HEASARC), which is a service of the Astrophysics Science Division at NASA/GSFC and the High Energy Astrophysics Division of the Smithsonian Astrophysical Observatory. The data used in this paper are public and available through the GSFC public archive at https://heasarc.gsfc.nasa.gov.
This work was made use of XRT and BAT data supplied by the UK $Swift$ Science Data Centre at
the University of Leicester\footnote{https://www.swift.ac.uk/swift\_ portal}.
%and MAXI data was provided by RIKEN, JAXA\footnote{http://maxi.riken.jp/mxondem}, and the MAXI team.

%\bibliographystyle{mnras}
\bibliographystyle{aa}


\end{document}